\documentclass[preprint,aps, showpacs,preprintnumbers,amsmath,amssymb,superscriptaddress]{revtex4-1}
\usepackage{amsmath,amsfonts,amssymb,graphics,graphicx,epsfig,color,times,indentfirst,layout,hyperref,subfigure,multirow}
\usepackage{hyperref}
\usepackage{ulem}
\hypersetup{linktocpage,,colorlinks=true}
\usepackage{threeparttable}
\newcommand{\dg}{^{\dagger}}

\begin{document}

\title{
M\"obius Kondo Insulators
}

\author{Po-Yao Chang} 
\email{pychang@physics.rutgers.edu}
\affiliation{Center for Materials Theory, Rutgers University, Piscataway, New Jersey, 08854, USA }

\author{Onur Erten}
\affiliation{Center for Materials Theory, Rutgers University, Piscataway, New Jersey, 08854, USA }
\affiliation{Max Planck Institute for the Physics of Complex Systems, Dresden, 01187, Germany}

\author{Piers Coleman}
\email{coleman@physics.rutgers.edu}
\affiliation{Center for Materials Theory, Rutgers University, Piscataway, New Jersey, 08854, USA }
\affiliation{Department of Physics, Royal Holloway, University of London, Egham, Surrey TW20 0EX, UK }

\begin{abstract}
{\bf Heavy fermion materials have recently attracted attention
for their potential to combine topological protection with strongly correlated
electron physics. To date, the ideas of topological
protection have been restricted to the heavy fermion
or ``Kondo'' insulators  with the simplest
point-group symmetries.  Here we argue that the presence of 
nonsymmorphic crystal symmetries in many heavy fermion
materials opens up a new family of topologically protected
heavy electron systems. 
Re-examination of archival resistivity measurements in nonsymorphic
heavy fermion insulators 
Ce$_3$Bi$_4$Pt$_3$ and CeNiSn 
reveals the presence of
low temperature conductivity plateau, making them 
candidate members of the new class of material. 
We illustrate our ideas with a specific model for CeNiSn, showing how 
glide symmetries generate surface states with a
novel M\" obius braiding that can be detected by ARPES or non-local
conductivity measurements.  
{One of the interesting effects of strong correlation, 
is the development  of partially localization or
``Kondo breakdown'' on the surfaces, 
which transforms M\"obius
surface states into quasi-one dimensional conductors,
with the potential for novel electronic phase transitions. }}
\end{abstract}

\maketitle

\noindent The discovery of topological phases of matter, initiated by
the pioneering works on quantum Hall states in
1980s\cite{Laughlin,Thouless,Haldane} has now evolved into the broad
notion of symmetry-protected topological states of matter. 
Heavy fermion or ``Kondo" insulators have recently 
emerged as a  particularly promising platform to
study the interplay between topological phases and strong electron
interactions\cite{Dzero, Dzero2012}. 
In topological Kondo insulators (TKIs), the strong
interaction between conduction electrons and local magnetic moments
leads to the formation of a narrow gap associated with the development
of Kondo singlets which screen the local moments\cite{Aeppli, Fisk,
Tsunetsugu, Riseborough}.  In SmB$_6$, the oldest known Kondo insulator, 
the existence of metallic surface states has been
demonstrated by transport experiments\cite{Wolgast, Kim, Kim2014,
Thomas} and angle-resolved photoemission spectroscopy
(ARPES)\cite{Neupane, Xu2013,Jiang2013}.  These results have
identified SmB$_6$ as a promising candidate for a TKI.

{
One of the important aspects of these materials is the interplay between
crystalline symmetry and topological order. To date, 
the main focus of interest in heavy fermion materials 
has been limited to the simplest crystalline symmetries.
In this work, we expand this notion to  a wider
class of heavy fermion materials in which unique topological features 
can arise from the combination of fractional translations and
by point group transformations known as nonsymmorphic symmetries.
Examples of 
Kondo insulators with non-symorphic symmetries include 
Ce$_3$Bi$_4$Pt$_3$\cite{Hundley,Cooley_PRB97}, and 
CeNiSn, CeRhSb, CeIrSb \cite{Hiess94,Sera97,Kawasaki_PRB2007}.
The observation of resistivity saturation at low temperatures 
in Ce$_3$Bi$_4$Pt$_3$ under pressure\cite{Cooley_PRB97}
and CeNiSn\cite{Slebarski} with Sb doping, closely resemble the
conductivity plateau of topological SmB$_6$, 
strongly suggesting that these nonsymmorphic Kondo insulators
are topologically nontrivial.
}


\begin{figure*}[htbp]
\centering
   \includegraphics[height=9. cm] {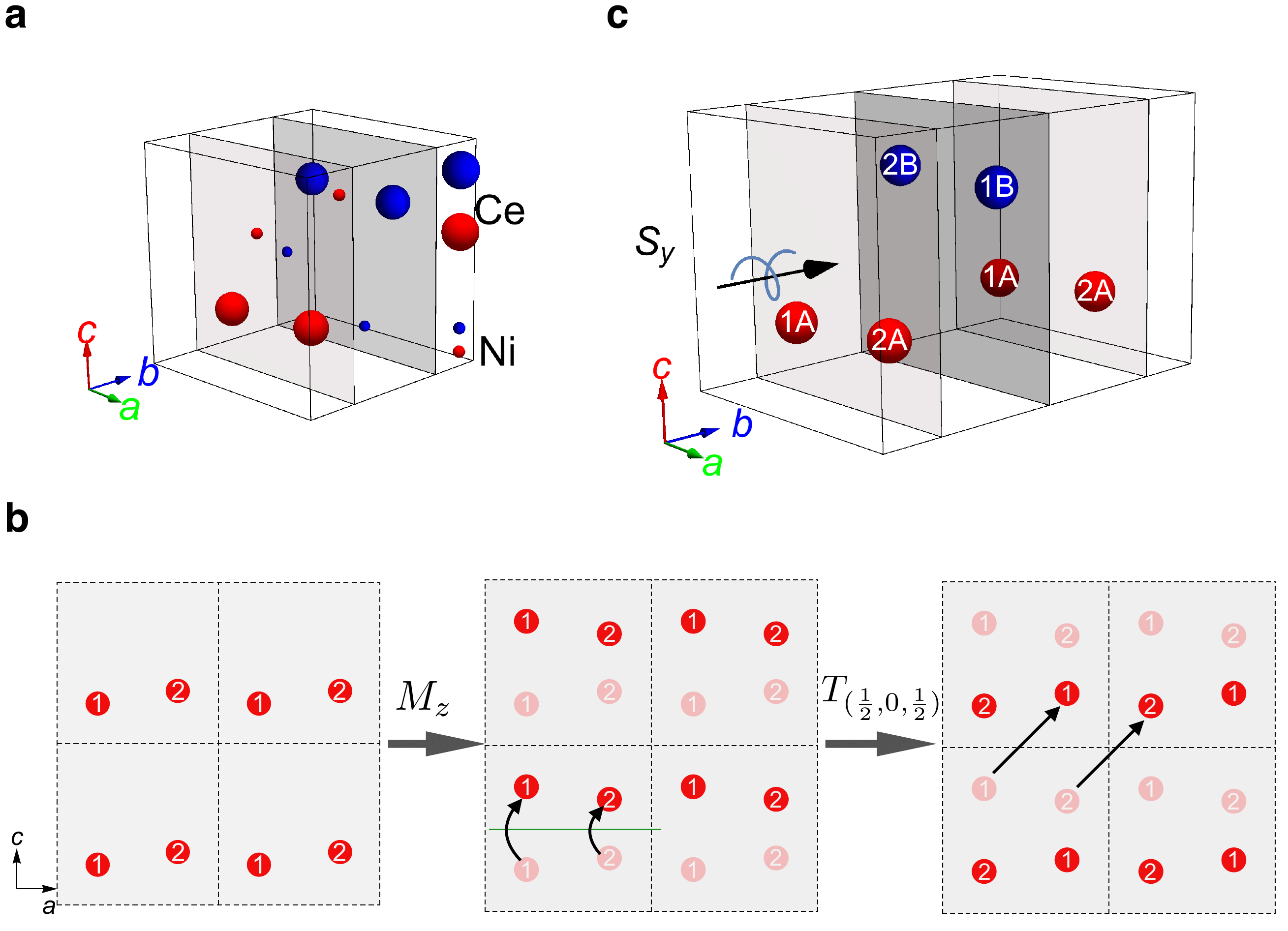}
\caption{{\bf Crystal structure of CeNiSn: } showing {\bf a} the
location of the Ce and Ni atoms, denoted by red and blue spheres. There are four equivalent Ce sites
$\{1A, 2A, 1B, 2B\}$ related by glide reflections and screw
rotations. The Ce and Ni atoms form zigzag chains 
which reside on two planes labeled $A$ (light gray) and $B$ (dark
gray), stacked along the $b$-axis.
{\bf b}, showing the glide reflection $G_z=T_{(1/2,0,1/2)}M_z $
which connects the Ce sites on different chains $G_z:(1A, 2A, 1B,
2B)\rightarrow (2A, 1A, 2B, 1B)$. {\bf c} showing the screw rotation
$S_y =T_{(0, 1/2, 0)}R_y^{\pi} $ which connects the chains in different planes $S_y: (1A, 2A, 1B, 2B)\rightarrow(1B, 2B, 1A, 2A)$.
}
\label{Fig: s1}
\end{figure*}


{
To illustrate the topological effects of 
nonsymmorphic symmetries in heavy fermion systems, here we study
CeNiSn as a representative member of this new new class of TKIs.} 
A central element of our theory is a tight-binding model of CeNiSn with all its salient symmetries.
The structure of CeNiSn belongs to nonsymmorphic space group No. 62 ($Pnma$),
containing three glide reflections, three screw rotations and
inversion symmetry, as shown in Fig. \ref{Fig: s1}.  
Glide reflections and screw rotations are nonsymmorphic symmetries,
which combine a point group operation (mirror or rotation) and a fractional lattice translation.  
We find that these symmetries permit
nonsymmorphic Kondo insulators to develop a protected surface state,
composed of two Dirac cones. Unlike regular topological insulators, in
which scattering between two surface Dirac cones can open a gap
without breaking time-reversal symmetry, the autonomy of these surface
states is stabilized by glide reflection and time-reversal symmetry
\cite{Shiozaki2015,Wang_arxiv2015}.  Moreover, nonsymmorphic symmetries give rise to
a momentum dependent twist that enables the surface states to be detached
from the bulk on the glide plane.  Following recent
studies\cite{Lu_2016, Fang2015, Shiozaki, Shiozaki2015, Wang_arxiv2015}, we
refer to these states as M\"obius-twisted surface states.  From the
bulk-boundary correspondence, we are able to define a $\mathbb{Z}_4$
topological invariant, and discuss the experimental signatures of such
a phase. 
{ One of the important effects that sets these topological
insulators apart from their weakly interacting counterparts, is the
possibility of breakdown of
the Kondo effect at the surface\cite{Alexandrov_PRL2015}.  
We find that this breakdown has a particularly dramatic effect on the 
M\"obius-twisted surface states, giving rise to 
quasi-one dimensional Fermi surfaces.
}

\section{Tight-binding Hamiltonian and nonsymmorphic symmetries}

We begin by constructing a tight-binding Hamiltonian for CeNiSn.
CeNiSn has an orthorhombic $\epsilon$-TiNiSi structure belonging to 
nonsymmorphic space group No. 62, $Pnma$,
which contains an inversion $P$, a screw rotation $S_y=T_{(0, 1/2,
0)}R^\pi_y$, and a glide reflection $G_z=T_{(1/2, 0, 1/2)}M_z $,
where $R^\pi_i$ denotes a $\pi$ rotation about the $i$-axis, 
$M_j$ refers to the mirror operation in the plane perpendicular to the
to the $j$-axis,
and $T_{(a,b,c)}$ is the translation operator along $a \hat{x} +b \hat{y}+c \hat{z}$.
There are four equivalent Ce sites in the unit cell which 
we label as $\{1A, 2A, 1B, 2B\}$ as shown in Fig \ref{Fig: s1}. 
The Cerium sites form zig-zag chains in the ac plane which are stacked 
along the c direction. Fig. \ref{Fig: s1}{\bf b} shows how glide reflection connects inter-chain 
sites $G_z:(1A, 2A, 1B, 2B)\rightarrow (2A, 1A, 2B, 1B)$.
These layers are then arranged in an alternating fashion along the b
direction; the alternating layers are related
related by the screw rotation $S_y:(1A, 2A, 1B, 2B)\rightarrow(1B, 2B, 1A, 2A)$
as shown in Fig. \ref{Fig: s1}{\bf c}.
In the following discussion, we re-scale the dimensions $a$, $b$ and $c$ of the unit cell to be unity. 
When applying the glide reflection and the screw rotation symmetries twice,
the system is shifted by a lattice translation, but the process also
involves a double reflection or $\pi$ rotation. The half-integer
character of the electrons means that  reflections or $\pi$-rotations
square to $-1$, and this additional factor means appears in the square
the square of glide reflection and screw rotations as follows:

\begin{align}
G_z^2&=T_{(1/2,0,1/2)}M_zT_{(1/2,0,1/2)}M_z\notag\\
&=T_{(1,0,0)} M_z^2 = -T_{(1,0,0)}\equiv -e^{-ik_x}.
\label{Eq: G2}
\end{align}
\begin{align}
S_y^2&=T_{(0,1/2,0)}R^\pi_yT_{(0,1/2,0)}R^\pi_y\notag\\
&=T_{(0,1,0)}[R^\pi_y]^2 = -T_{(0,1,0)}\equiv -e^{-ik_y}.
\label{Eq: S2}
\end{align}

Band structure calculations\cite{Yanase92, Hammond95, Hiess97} indicate that 
the relevant orbitals near the chemical potential derive from the Ce $4f$-electrons and Ni $3d$-electrons. We
now construct a simplified model involving these two sets of
orbitals. A key ingredient of our model is the hybridization between
the $f$ and $d$ states which involves 
the transfer of one unit of angular momentum from spin, to orbital
angular momentum. As a result, the 
hybridization develops a p-wave form-factor\cite{Alexandrov_PRL2015},
and can be modelled by a simpler model 
of spin-orbit $p$ orbitals hybridizing with $s$-wave conduction electrons. 
We project the Wannier states of these two sets of orbitals onto the
common sites of the Cerium atoms. 
The resulting 
tight-binding Hamiltonian has the structure
\begin{align}
&\mathcal{H}^{}({\bf k})=
\left(\begin{array}{cc} \mathcal{H}^c({\bf k}) & V({\bf k}) \\ V^\dagger({\bf k}) & \mathcal{H}^f({\bf k})\end{array}\right),  
\label{Eq: Ham_tot}
\end{align}
where $V({\bf k})$ is the hybridization matrix, $\mathcal{H}^{c}$ and $\mathcal{H}^{f}$
are the nearest hopping matrices for the conduction and 
$f$-electrons respectively. The detailed structure of this
Hamiltonian, which respects the full nonsymmorphic symmetries
of the lattice, is provided in the methods. 

\begin{figure*}[htbp]
\centering
\includegraphics[width=15 cm] {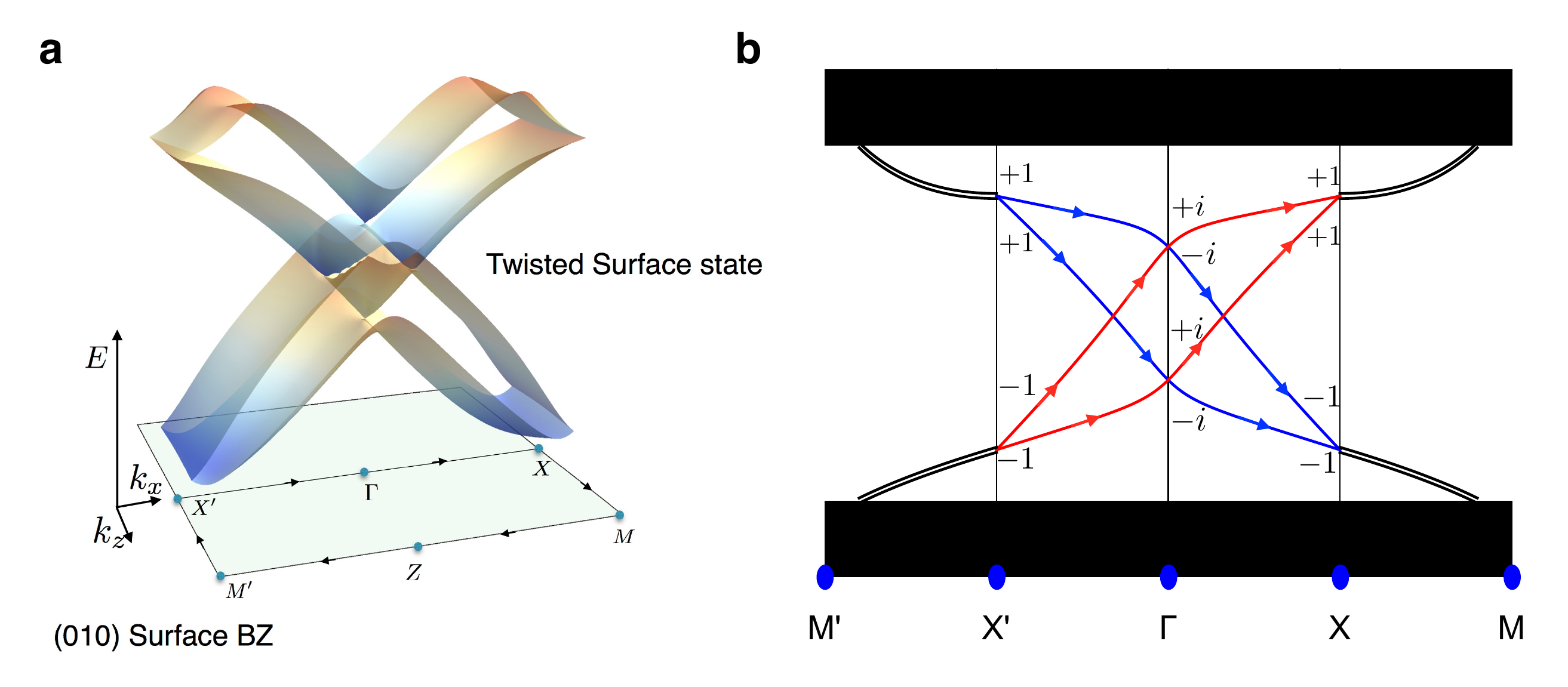}
\caption{{\bf A pair of M\"obius twisted surface states on the $(010)$
surface. }
{\bf a}. Three dimensional plot of the dispersion on
 the $(010)$ surface.  The two 
M\"obius twisted surface states are detached
from the bulk along the glide line ($X'\Gamma X$).  {\bf b}. Schematic
showing dispersion along the glide line $X'\Gamma
X$, where red and blue lines correspond to states
with positive and negative glide eigenvalues $g_{\pm}$, respectively.
Along $XM$ $(X'M')$ the surface states are doubly degenerate due to a
combination of time-reversal and glide symmetry, 
$\mathcal{T}\mathcal{G}_z$. The imaginary glide eigenvalues
$\pm i$
at the $\Gamma$ point force the members of each Kramers pair to belong
to different glide sectors, but at the $X (X')$ points 
the real glide eigenvalues $\pm 1$
mean that the members of a Kramers pair are in the same glide sector.
The connectivity between
Kramers pairs at the $\Gamma$ and $X^{(')}$ points gives rise to the
M\" obius character, for 
if one starts at $X'$ and follows the red loop to $X$, one has to pass
a second time around the loop on the blue line, before one returns to
the origin. 
}
\label{Fig: F2}
\end{figure*}

\begin{figure*}[htbp]
\centering
\includegraphics[height=7.5 cm] {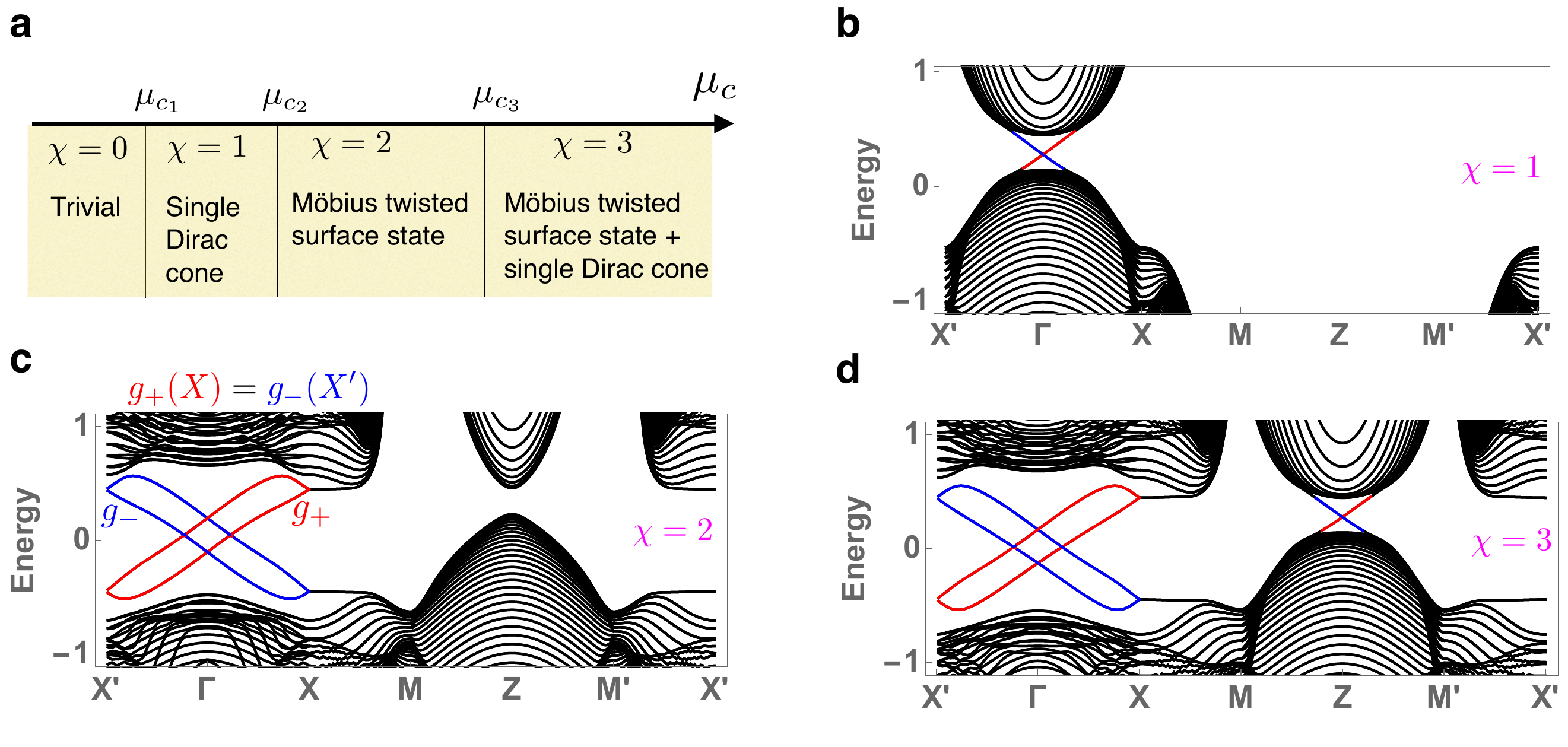}
\caption{{\bf $\mathbb{Z}_4$ topological invariants and the corresponding surface states: }
{\bf a}, Four phases as the function of the chemical potential of
conduction electrons $\mu_c$, with
$(\mu_{c_1},\mu_{c_2},\mu_{c_3})=(-73,-60.5,-33)$ and the parameters
used in equation (\ref{Eq: Ham_tot}) are $(\alpha, \beta,
\gamma,a,b)=(0.525,0.525,01,0.5,0.5)$,
$(t_x^c,t_y^c,t_z^c)=(5,-25,-10) =-20(t_x^f,t_y^f,t_z^f)$, and
$\mu_c=-20 \mu_f$.  {\bf b-d}, The energy spectra are computed in a
$(010)$ slab geometry for $\chi=1,2,3$, respectively.  Red and blue
lines are the energy dispersion of surface states with positive and
negative glide eigenvalues $g_{\pm}$, respectively.  A M\"obius
twisted surface is located at the glide plane ($X'\Gamma X$) shown in
{\bf c} and {\bf d}.  This M\"obius twisted surface is composed
by two Dirac cones.  Due to Kramers pairs at $X$ point have the same
glide eigenvalue and $g_+(X)=g_-(X')$, this M\"obius twisted
surface is detached from the bulk at the glide plane.
}
\label{Fig: F3}
\end{figure*}

\section{Topological surface states with a  M\"obius twist}

{One symmetry-preserving surface which respects to the glide reflection
$\mathcal{G}_{z}$ is the $(010)$ surface.}
 This surface is perpendicular to the glide plane {($xy$ plane)}
 and is also invariant under lattice translations
parallel to the surface.  The surface energy dispersion as a function of
$(k_x, k_z)$ is computed by diagonalizing the Hamiltonian in a
$(010)$ slab geometry.  The corresponding surface Brillouin zone (BZ)
is shown in Fig. \ref{Fig: F2}{\bf a}.  The glide lines on the surface
BZ are the set of glide reflection invariant momenta, which are at
$k_z=0$ (path $X'\Gamma X$) and $k_z=\pi$ (path $M'ZM$). Along these
lines, the Hamiltonian from equation (\ref{Eq: Ham_tot}) commutes
with $\mathcal{G}_z$ and can be block diagonalized into two sectors with
two eigenvalues for ${\mathcal{G}_z}$, 
$g_{\pm} (k_x)= \pm i e^{- i k_x/2}$ along the glide lines.

On the glide lines along 
$X'\Gamma X$ or $M'ZM$, a pair of surface Dirac cones is
stabilized by the glide reflection and time-reversal symmetry.  To
demonstrate this state (see Fig.  \ref{Fig: F2}{\bf b}), 
we focus on path $X'\Gamma X$.  At the $X(X')$ point, the glide
eigenvalues are real ($\pm 1$), which implies that the members of each
Kramers pair derive from the same glide sector, i.e., the glide
eigenvalues for two Kramers pairs are $(+1,+1)$ and $(-1,-1)$.  By
contrast, at the $\Gamma$ point, the glide eigenvalues are imaginary
($\pm i$), so time reversal inverts the glide eigenvalue, which
indicates that the members of each Kramers pair come from opposite
glide sectors, i.e., the glide eigenvalues for two Kramers pairs are
both $(+i,-i)$.  When we connect two Kramers pairs at $\Gamma$ point
to two Kramers pairs at $X(X')$ point we obtain the hourglass
structure of this surface state\cite{Wang_arxiv2015}, which contains
two Dirac cones at the $\Gamma$ point (Fig. \ref{Fig: F2}{\bf b}).  This
surface state contains a M\"obius twist, for if we follow the arrow
from Fig. \ref{Fig: F2}{\bf b} along the loop $X'\Gamma X$, we need go
around the loop twice: once on a red and once on a blue branch, before
returning to the origin.  Due to this unusual connectivity, the
surface state can be detached from the bulk along the loop $X'\Gamma
X$.  Fig.  \ref{Fig: F2}{\bf a} displays the result of a band-calculation on a
strip, showing the M\" obius-twisted character.

The presence or absence of a M\" obius-twisted surface state on the (010) surface
defines a $\mathbb{Z}_{2}$ variable. When we combine this with the additional
$\mathbb{Z}_{2}$ variable associated with the 
possibility of forming a strong topological insulator, by introducing 
an additional odd number of Dirac cones on every surface, 
we see that the combination of time-reversal and non-symmorphic
symmetries gives rise to a $\mathbb{Z}_{4}$ topological invariant
$\chi$, for which we can construct a corresponding $\mathbb{Z}_{4}$ index
(see Supplementary Information and also Ref. \cite{Shiozaki2015}), as
shown in  Fig. \ref{Fig: F3}{\bf a}. 
$\chi=0$ corresponds to a trivial insulator with no gapless surface
states.  
$\chi=1$ corresponds to a strong topological
insulating phase with one single surface Dirac cone (Fig.  \ref{Fig: F3}{\bf b}).
$\chi=2$ corresponds to a nonsymmorphic topological insulator
with a M\"obius twisted surface state along $X'\Gamma X$
$(M'ZM)$ path (Fig.  \ref{Fig: F3}{\bf c}), while
$\chi=3$ corresponds to a strong topological
insulating phase with three surface Dirac cones (Fig.  \ref{Fig: F3}{\bf d}).


In our model calculations, we also observe a double Dirac cone like surface state on $(001)$ surface,
where the crossings are located at $(k_x,k_y)=(\pm k_0, 0)$.
However at the mirror plane $k_y=0$, this surface state is gapped and
is not protected by mirror symmetry $\mathcal{M}_y$ and time-reversal
symmetry $\mathcal{T}$, so this state will likely be absent in the
real material. 


 \begin{figure*}[htbp]
\centering
   \includegraphics[height=5. cm] {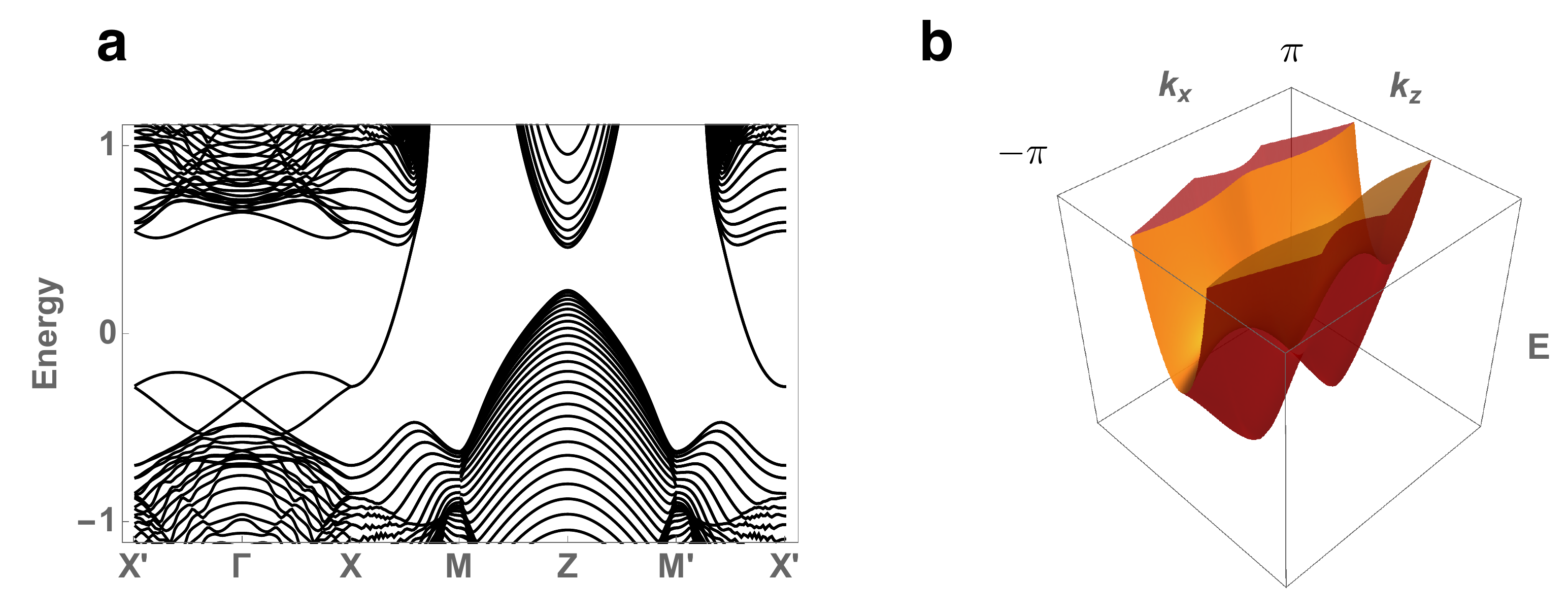}
\caption{
{\bf 
{\bf a,} Showing the effect of Kondo breakdown on the
dispersion of the M\"obius-twisted surface state, causing the Dirac points
to sink into the valence band.
{\bf b,} Three dimensional plot showing the quasi-one dimensional 
dispersion of M\"obius-twisted surface state  in the presence of Kondo
breakdown. }}
\label{Fig: s4}
\end{figure*}

\section{Discussion}

We have shown that CeNiSn and Kondo insulators with nonsymmorphic
symmetries have the potential to form a new class of topological Kondo
insulators with unusual surface states.  CeNiSn is of course a
low-carrier density metal, with a small Fermi surface derived from an indirect
band-gap closure or a lightly doped conduction band\cite{Terashima02,Paschen:2016}, but such small
bulk Fermi surfaces are readily localized by disorder or substitution.
This is the likely explanation of the observation of a resistivity
plateau below 10K in antimony-doped
CeNiSn$_{1-x}$Sb$_{x}$\cite{Slebarski}, where the observation of a
resistivity plateau most likely derives from a metallic topological
surface states, as in the case of SmB$_6$. Moreover the V-shape
density of states deduced from NMR experiments and point-contact
spectroscopy\cite{Ekino,
Nakamura96,Nakamura_JPSJ1994} can be 
accounted for as a signatures of Dirac cone surface states.  The large magnetoresistance for fields
perpendicular to the $a$ axis\cite{Terashima02} may be a consequence of metallic
surface in the $(010)$ plane, combined with an insulating surface
on the $(100)$ and $(001)$ planes.
{ Although the above discussion ahs focussed on CeNiSn, we note
that the nonsymmorphic Kondo insulator Ce$_{3}$Bi$_{4}$Pt$_{3}$
also displays a reistivity plateau. Remarkably, in the presence of
pressure, the  resistivity plateau persists up to 100K\cite{Cooley_PRB97} [See supplementary material]. 
These early experimental results provide strong circumstantial 
support for the topological nature of these nonsymmorphic Kondo insulators
and provide a strong motivation for further detailed investigation.
}

The confirmation of these ideas requires a direct probe of
M\" obius conducting surface states, either by ARPES measurements, or by 
non-local\cite{Wolgast} or sample thickness dependent
transport measurements\cite{Syers}.  One of the interesting challenges is 
to delineate M\"obius surface states from 
single Dirac surface states. According to
Ref. \cite{Nakajima}, one-dimensional chiral edge modes from
ferromagnetic domain walls on the surface of SmB$_6$ have been
observed with quantized $e^2/h$ conductance.  
If an analogous ferromagnetic order  emerges on a nonsymmorphic
surface (induced by field or magnetic impurities), any 
M\"obius state present will become gapped.  The Landau levels of a M\"obius
surface state will contribute to quantized conductance $2
(n+1/2) e^2/h$, where $n$ depends on the surface chemical
potential. When the Fermi energy is in the gap, the conductance only
comes from the lowest Landau level with $n=0$.  The transport
measurement presumably comes from both top and bottom surface that
give rise to the total conductance $2 e^2/h$, twice that
observed in SmB$_6$.  Another way to detect
twisted surface states is from Hall bar measurements. In contrast with
the usual topological insulators, where the
quantum Hall conductivity $\sigma_{H}$ switches from $-e^2/h$ to
$e^2/h$ on gating, the twisted surface state will generate three
values of quantized Hall conductivity.  When the gate voltage is
below(above) two cones, we have $\sigma_H=-e^2/h$($e^2/h$).  Whereas
the gate voltage is in between two cones, the Hall conductance will
vanish, $\sigma_H=0$.

{ The strong electron correlations in Kondo insulators opens up
the possibility of many interesting phenomena, absent in their weakly
interacting counterparts. Here, a particularly important phenomenon is
the possibility of a surface breakdown of the Kondo
effect\cite{Alexandrov_PRL2015}.  Surface Kondo breakdown is based on
the observation that the reduced coordination of the rare earth ions
at the surface causes a reduction of in the surface Kondo
temperature. In principle, competing instabilities such as magnetism
can now be activated on the surface. The breakdown of the the Kondo
singlets at the surface liberates a large number of conduction
electrons which then dope the topological surface states. We have
examined  the
effects of surface Kondo breakdown in M\" obius surface states by
recomputing the surface spectrum in the absence of the surface
f-states. These calculations show 
that the Kondo breakdown causes the Dirac points in 
M\"obius surface states to sink into the valence
band as shown in Fig. \ref{Fig: s4}{\bf a} which then generates large
Fermi surfaces [see Fig. \ref{Fig: s4}{\bf b}].  The detached
nature of the M\"obius surface state causes the resulting 
Fermi surfaces to become quasi
one-dimensional along $k_x$ direction.  The interaction of the
partially unscreened surface local moments with
these quasi one-dimensional Fermi surfaces is expected to lead to a
wide variety of surface electronic instabilities, 
including unconventional
superconductivity\cite{Bennemann_Book} and charge or spin density wave
instabilities.  }

Another interesting future direction is the possibility of nonsymmorphic topological
superconductors. Promising candidates are UCoGe and URhGe, which share the
same space group as CeNiSn\cite{Huy_PRL2007, Aoki_Nature2001}. These
materials exhibit spin-triplet superconductivity in coexistence 
with ferromagnetism. The topological classification of such
superconductors is an intriguing future prospect.

\section{Methods}

In the momentum space, the tight-binding Hamiltonian 
is $H = \sum_{\bf k}\Psi({\bf k})^\dagger \mathcal{H}^{} ({\bf
k})\Psi({\bf k})$,
where $\Psi({\bf k})$ is a  sixteen component spinor, $\Psi({\bf k})=(\Psi_c({\bf k}),\Psi_f({\bf k}))^{\mathrm{T}}$ with
\begin{align} 
\Psi_{c}({\bf k})=&(c_{1A\uparrow }({\bf k}),c_{1A\downarrow}({\bf k}),c_{2A\uparrow }({\bf k}),c_{2A\downarrow}({\bf k}),  \notag\\
&c_{1B\uparrow }({\bf k}),c_{1B\downarrow}({\bf k}),c_{2B\uparrow }({\bf k}),c_{2B\downarrow}({\bf k})),   \notag \\
\Psi_{f}({\bf k})=&(f_{1A\uparrow }({\bf k}),f_{1A\downarrow}({\bf k}),f_{2A\uparrow }({\bf k}),f_{2A\downarrow}({\bf k}), \notag\\
&f_{1B\uparrow }({\bf k}),f_{1B\downarrow}({\bf k}), f_{2B\uparrow }({\bf k}),f_{2B\downarrow}({\bf k})),   \notag
\end{align}
and
\begin{align}
&\mathcal{H}^{}({\bf k})=
\left(\begin{array}{cc} \mathcal{H}^c({\bf k}) & V({\bf k}) \\ V^\dagger({\bf k}) & \mathcal{H}^f({\bf k})\end{array}\right),  
\label{Eq: Ham_tot}
\end{align}
where $V({\bf k})$ is the hybridization matrix, $\mathcal{H}^{c}$ and $\mathcal{H}^{f}$
are the nearest hopping matrices for conduction and $f$-electrons, respectively.
In order to simplify our calculation, we introduce four sets of Pauli matrices: 
$\{\sigma_i\}$ acts on the spin basis; $\{\lambda_i\}$ acts on the basis of conduction electrons and $f$-electrons;
$\{\tau_i\}$ acts on the basis of the atom labels $1$ and $2$; $\{\rho_i\}$ acts on the basis of the layer labels $A$ and $B$.

From equation (\ref{Eq: Ham_tot}), the hybridization matrix has the form
\begin{align}
V({\bf k})=
\left(\begin{array}{cc} V_A({\bf k}) & V_{AB}({\bf k}) \\ V_{BA}({\bf k}) & V_B({\bf k})\end{array}\right),    \notag
\label{Eq: Ham_tot_1}
\end{align}
where
\begin{align}
&V_A({\bf k})  =
\left(\begin{array}{cc}  2 i t_2 \sin k_z & t_1+\sigma_3 t_1 \sigma_3 e^{-i k_x} 
\\ -t_1-\sigma_3 t_1 \sigma_3 e^{i k_x} & 2i t_2 \sin k_z 
\end{array}\right)= - V_B^{} ({-\bf k}),   \notag \\
&V_{AB}^{} ({\bf k}) =
\left(\begin{array}{cc}   &  t_3-\sigma_2 t_3 \sigma_2 e^{-ik_y} 
 \\  t_4- \sigma_2 t_4 \sigma_2e^{-ik_y}  & \end{array}\right), \notag\\
&V_{BA}^{} ({\bf k}) =
\left(\begin{array}{cc} & \sigma_2 t_3 \sigma_2 e^{ik_y} -  t_3  
\\ \sigma_2 t_4 \sigma_2 e^{ik_y}- t_4  &\end{array}\right),   \notag
\end{align}
with $t_{1} = i (\alpha \sigma_1+\beta \sigma_3)$, $t_{2} = i \gamma \sigma_3$,
$t_3= i (a \sigma_2 +b \sigma_3)$, and $t_4= i (a \sigma_2 -b \sigma_3)$.

The nearest hopping matrices 
 for conduction electrons and $f$-electrons are
\begin{align}
\mathcal{H}^{l} ({\bf k})=& (2 t^l_z \cos k_z - \mu^l) + 2 t^l_x \cos \frac{k_x}{2} ( \cos \frac{k_x}{2} \tau_1 + \sin \frac{k_x}{2} \tau_2 \rho_3) \notag\\
&+2 t^l_y \cos \frac{k_y}{2} ( \cos \frac{k_y}{2} \tau_1 \rho_1 + \sin \frac{k_y}{2} \tau_1 \rho_2),    \notag
\end{align}
where $l = c,f$, $t^l_i$ are the hopping amplitudes along $i$-direction, 
and $\mu^l$ are the bare energies of the
isolated conduction electrons  and $f$-electrons.
In the Supplementary Information we perform the construction of this tight-binding Hamiltonian in detail.

We write down the matrix representations of symmetries as follows:
\begin{enumerate}
  \item  Time-reversal symmetry, $\mathcal{T}^{-1} \mathcal{H} ({\bf k}) \mathcal{T} =  \mathcal{H} (-{\bf k})$,
where $ \mathcal{T}= i \sigma_2 \mathcal{K}$ with $\mathcal{K}$ being the complex conjugation operator.
  \item Inversion symmetry, $\mathcal{P}^{-1} \mathcal{H} ({\bf k}) \mathcal{P} =  \mathcal{H} (-{\bf k})$,
where $\mathcal{P}=\lambda_3 \rho_1$.
  \item Glide reflection symmetry $\mathcal{G}_z$, 
${\mathcal{G}_z}^{-1} \mathcal{H} (k_x,k_y,k_z) {\mathcal{G}_z} =  \mathcal{H} (k_x,k_y,-k_z)$,
where 
\begin{align}
{\mathcal{G}_z}({\bf k})
= -ie^{-i\frac{k_x}{2}}\sigma_3(\cos \frac{k_x}{2} \tau_1 + \sin \frac{k_x}{2} \tau_2 \rho_3) \lambda_3.  \notag
\end{align}
\item Screw rotation symmetry $\mathcal{S}_y$, 
${\mathcal{S}_y}^{-1} \mathcal{H} (k_x,k_y,k_z){\mathcal{S}_y} =  \mathcal{H} (-k_x,k_y,-k_z)$,
where
\begin{align}
{\mathcal{S}_y}({\bf k}) = -ie^{-i\frac{k_y}{2}}\sigma_2(\cos \frac{k_y}{2} \rho_1+\sin \frac{k_y}{2} \rho_2).  \notag
\end{align}

{
\item Mirror symmetry $\mathcal{M}_y=\mathcal{S}_y\mathcal{P}$, 
${\mathcal{M}_y}^{-1} \mathcal{H} (k_x,k_y,k_z){\mathcal{M}_y} =  \mathcal{H} (k_x,-k_y,k_z)$,
where
\begin{align}
{\mathcal{M}_y}({\bf k}) = -ie^{-i\frac{k_y}{2}}\sigma_2(\cos \frac{k_y}{2} \rho_0- i \sin \frac{k_y}{2} \rho_3)\lambda_3.  \notag
\end{align}}
\end{enumerate}

In the spin-orbit coupled systems, reflection and $\pi$ rotation square to $-1$. 
We have ${\mathcal{G}_z}({\bf k})^2 = -e^{ - i k_x}$, ${\mathcal{S}_y}({\bf k})^2 = -e^{ - i k_y}$, and ${\mathcal{M}_y}(k_y=0,\pi)^2 = -1$.

\section{Acknowledgments}

The authors would like to thank Silke Paschen and Tyrel McQueen for
discussions about CeNiSn. 
This work was supported by the Rutgers
Center for Materials Theory group postdoc grant (Po-Yao Chang), US
National Science Foundation grant 
grant DMR-1309929 (Onur Erten) and US Department of Energy grant 
DE-FG02-99ER45790 (Piers Coleman).

\section{Author contributions}
All authors performed the calculations, discussed the results and prepared the manuscript.

\section{Competing financial interests}
The authors declare no competing financial interests.

\bibliographystyle{naturemag}

\newpage

\section{
M\"obius Kondo Insulators: Supplementary Material
}

\maketitle

\subsection{A tight-binding model based on nonsymmorphic symmetries}
\label{App: Ham}
Here we derive the tight-binding Hamiltonian for CeNiSn using the
following procedure:
\begin{enumerate}
\item We define the symmetry operations on the fermionic operators.
\item We then construct the Hamiltonian by writing down the all the nearest
neighbor hopping and hybridization terms which respect the
nonsymmorphic symmetries. 
\end{enumerate}
In order to simplify our calculation, we introduce four sets of Pauli matrices: 
$\{\sigma_i\}$ acts on the spin basis; $\{\lambda_i\}$ acts on the basis of conduction electrons and $f$-electrons;
$\{\tau_i\}$ acts on the basis of the atom labels $1$ and $2$; $\{\rho_i\}$ acts on the basis of the layer labels $A$ and $B$.

In CeNiSn, $4f$-electrons of Ce atoms hybridize with
$3d$-electrons of Ni atoms. Since the total angular momentum difference
between these two states is one, 
we orbitally ``downfold'' the tight-binding model, replacing it by an
equivalent model, reducing the total angular momentum $J$ 
of each band by two units.  The resulting model involves the 
hybridization of spin-orbit coupled
p-electrons with s-electrons 
The highly localized states are then modelled as 
$p$-wave Kramers doublets,
\begin{align}
&f^{\dagger}_{\uparrow} |0 \rangle\equiv | p_{x\downarrow} + i p_{y\downarrow} +  p_{z\uparrow}  \rangle    \notag\\
&f^{\dagger}_{\downarrow} |0 \rangle\equiv  | p_{x\uparrow} - i p_{y\uparrow} -  p_{z\downarrow}  \rangle. 
\end{align}
while the mobile conduction electrons are $s$-wave,
\begin{align}
c^{\dagger}_{\uparrow/\downarrow} |0 \rangle\equiv  | s_{\uparrow/\downarrow} \rangle.
\end{align}
For further simplicity, we project the conduction electron Wannier states onto the sites of $f$-electrons.

\subsection{Symmetries:}
Here, we describe the actions of the various 
symmetry operations in  space group No. 62 ($Pnma$) on the Fermi
operators. This group contains three screw rotations $S_{x,y,z}$, 
three glide reflections $G_{x,y,z}$ and an inversion $P$.
In fact, this set of six operators can all be generated from 
$G_{z}$ $S_{y}$, in combination with inversion and translation
operators, so it is sufficient for us to focus on these two
non-symmorphic operators. 

\subsubsection{Glide reflection $G_z: (X, Y, Z) \to (X+1/2, Y, -Z+1/2)$}
The glide reflection $G_z$ transforms the $f$-electrons and conduction
electrons as follows
\begin{align}
&G_z^{-1} f^\dagger_{1L\sigma} ({\bf x}_{j}) G_z = 
- (i \sigma_3)_{\sigma\sigma'}  f^\dagger_{2L\sigma'} ({\bf x}_{j}+ c), 
\notag\\
&G_z^{-1} f^\dagger_{2L\sigma} ({\bf x}_{j}) G_z =
- (i\sigma_3)_{\sigma\sigma'}  f^\dagger_{1L\sigma'} ({\bf x}_{j}+ a+ c),
\end{align}
where $\vec{x}_{j}$ is the co-ordinate of the unit cell, 
$ (L\in [A,B]) $ and 
$1$ and $2$ are the site and layer indices 
of atoms within one unit cell, 
$a$
and $c$ are unit vectors in the $x$ and $z$ directions,
respectively, while $\sigma  $ is the spin index.
The glide reflection acts in a similar way on the conduction electrons
\begin{align}
&G_z^{-1} c^\dagger_{1L\sigma} ({\bf x}_{j}) G_z =  
(i \sigma_3)_{\sigma\sigma'}  c^\dagger_{2L\sigma'} ({\bf x}_{j}+c), 
\notag\\
&G_z^{-1} c^\dagger_{2L\sigma} (\vec{x}_{j}) G_z = 
(i \sigma_3)_{\sigma\sigma'}  c^\dagger_{1L\sigma'} ({\bf x}_{j}+a+c). 
\end{align}

\subsubsection{Screw rotation $S_y: (X,Y,Z) \to (-X,Y+1/2, -Z)$}
The screw rotation $S_y$ transforms the $f$-electrons and conduction
electrons
as follows: 
\begin{align}
&S_y^{-1} f^\dagger_{1(2)A\sigma}({\bf x}_j) S_y = i [\sigma_2]_{\sigma\sigma'}  f^\dagger_{1(2)B\sigma'}({\bf x}_j), \notag\\
&S_y^{-1} f^\dagger_{1(2)B\sigma}({\bf x}_j) S_y = i  [\sigma_2]_{\sigma\sigma'}  f^\dagger_{1(2)A\sigma'}({\bf x}_{j}+b),
\end{align}
and
\begin{align}
&S_y^{-1} c^\dagger_{1(2)A\sigma}({\bf x}_j) S_y = i  [\sigma_2]_{\sigma\sigma'}  c^\dagger_{1(2)B\sigma'}({\bf x}_j), \notag\\
&S_y^{-1} c^\dagger_{1(2)B\sigma}({\bf x}_j) S_y =i   [\sigma_2]_{\sigma\sigma'}  c^\dagger_{1(2)A\sigma'}({\bf x}_j+b), 
\end{align}
where $b$ is the unit vector in $y$ direction.

\subsubsection{Inversion $P: (X, Y, Z): \to (-X, -Y ,-Z)$}
The inversion $P$ transforms the $f$-electrons and conduction electrons in the following way,
\begin{align}
&P^{-1} f^\dagger_{1(2)A\sigma}({\bf x}_j) {P} = - f^\dagger_{1(2)B\sigma}({\bf x}_j), \notag\\
&P^{-1} f^\dagger_{1(2)B\sigma}({\bf x}_j) {P} =-f^\dagger_{1(2)A\sigma}({\bf x}_{j}).
\end{align}
And
\begin{align}
&{P}^{-1} c^\dagger_{1(2)A\sigma}({\bf x}_j) {P} = c^\dagger_{1(2)B\sigma}({\bf x}_j), \notag\\
&{P}^{-1} c^\dagger_{1(2)B\sigma}({\bf x}_j) {P} = c^\dagger_{1(2)A\sigma}({\bf x}_{j}).
\end{align}

\subsubsection{Other 
symmetries:}\label{othersym}

By combining inversion, translations, and the above two
nonsymmorphic symmetries we can obtain all the remaining
symmetries.  
Thus $G_x= T_{(1,0,1)}S_yG_z$,  and $G_y:(X,Y,Z) \to (X,-Y+1/2, Z)$, since
\begin{align}
(X, Y, Z) &\stackrel{G_z}{\longrightarrow} (X+1/2, Y, -Z+1/2)
 \stackrel{S_y}{\longrightarrow} (-X-1/2, Y+1/2, Z-1/2) \notag\\
 &\stackrel{T_{(1,0,1)}}{\longrightarrow} (-X+1/2, Y+1/2, Z+1/2),
\end{align}
and likewise, 
\begin{align}
(X, Y, Z) &\stackrel{S_y}{\longrightarrow} (-X, Y+1/2, -Z)
 \stackrel{P}{\longrightarrow} (X, -Y-1/2, Z) \notag\\
 &\stackrel{T_{(0,1,0)}}{\longrightarrow} (X, -Y+1/2, Z).
\end{align} 

In a similar fashion, we obtain 
$S_x= T_{(1,1,1)}PG_x$ and  $S_{z}=T_{(1,0,1)} P G_{z}$, 
since 
 \begin{align}
(X, Y, Z) &\stackrel{G_x}{\longrightarrow} (-X+1/2, Y+1/2, Z+1/2)
 \stackrel{P}{\longrightarrow} (X-1/2, -Y-1/2,-Z-1/2) \notag\\
 &\stackrel{T_{(1,1,1)}}{\longrightarrow} (X+1/2,-Y+1/2, Z+1/2).
\end{align} 
and 
\begin{align}
(X, Y, Z) &\stackrel{G_z}{\longrightarrow} (X+1/2, Y, -Z+1/2)
 \stackrel{P}{\longrightarrow} (-X-1/2, -Y, Z-1/2) \notag\\
 &\stackrel{T_{(1,0,1)}}{\longrightarrow} (-X+1/2,-Y, Z+1/2).
\end{align} 

\subsection{Construction of the Hamitonian}
\subsubsection{Hopping terms}
\begin{figure}[htbp]
\centering
\includegraphics[height=6cm] {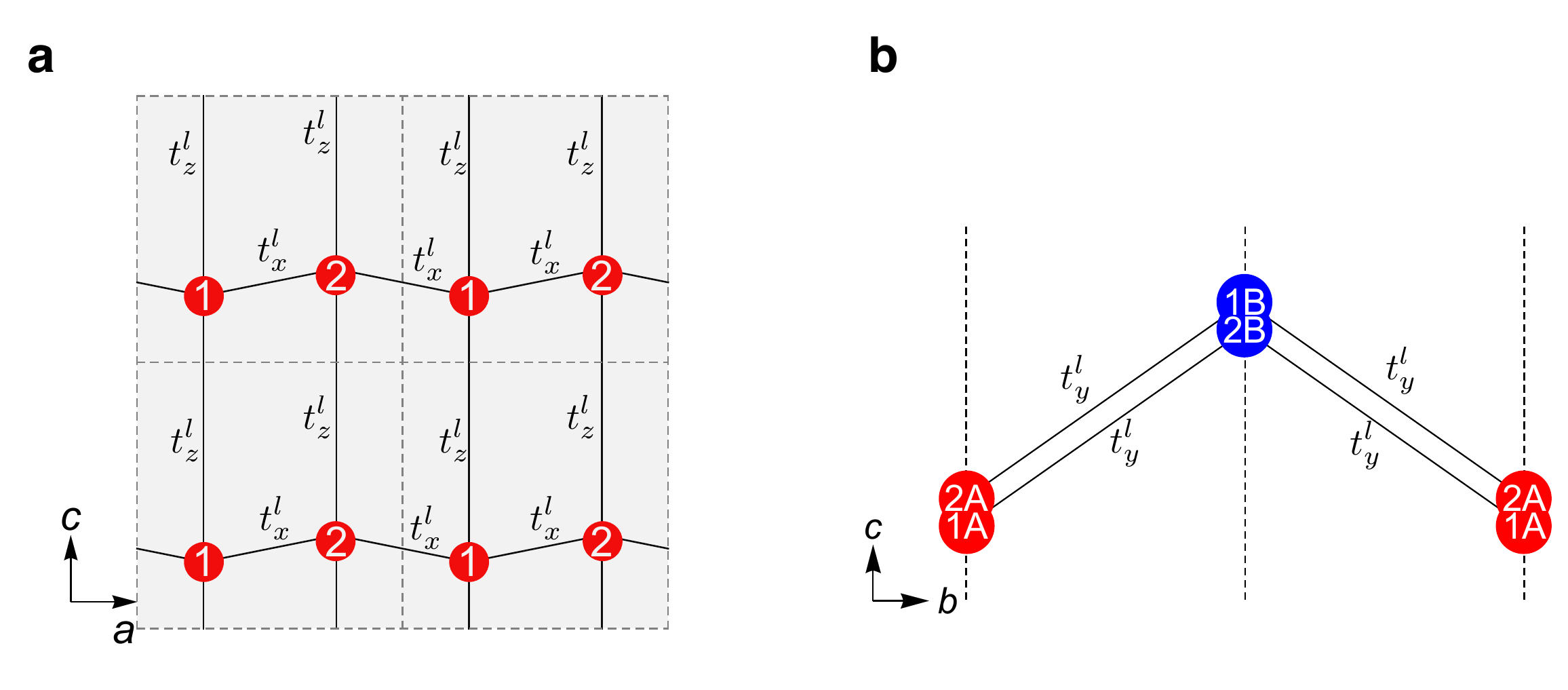}
\caption{{\bf Hopping Hamiltonian:}
{\bf a}, The hopping elements $t^l_x$ and  $t^l_z$ with $l=c,f$on the $ac$ plane.
{\bf b}, The hopping elements $t^l_y$ with $l=c,f$ on the $bc$ plane.}
\label{Fig: SM1}
\end{figure}

Now we consider the nearest neighbor hopping terms in the Hamiltonian [see Fig. \ref{Fig: SM1}].
\begin{align}
H^{\rm hopping}=& \sum_{j, \sigma,l=c,f}t^l_x  l^{\dagger}_{1A\sigma}({\bf x}_j)l_{2A\sigma}({\bf x}_j)
+t^l_x   l^\dagger_{2A\sigma}({\bf x}_j)l_{1A\sigma}({\bf x}_j+a) 
+t^l_z   l^\dagger_{1A\sigma}({\bf x}_j)l_{1A\sigma}({\bf x}_j+c)\notag\\  
&+t^l_z   l^\dagger_{1B\sigma}({\bf x}_j)l_{1B\sigma}({\bf x}_j+c)  
+t^l_y l^\dagger_{1A\sigma}({\bf x}_j)l_{2B\sigma}({\bf x}_j)  
+t^l_y l^\dagger_{2A\sigma}({\bf x}_j)l_{1B\sigma}({\bf x}_j) \notag\\
&+t^l_y l^\dagger_{1B\sigma}({\bf x}_j)l_{2A\sigma}({\bf x}_j+b)
+t^l_y l^\dagger_{2B\sigma}({\bf x}_j)l_{1A\sigma}({\bf x}_j+b) \notag\\
&-\sum_{\alpha=1,2;L=A,B;j;\sigma;l=c,f} \mu^l l^{\dagger}_{\alpha L\sigma}({\bf x}_j)l_{\alpha L \sigma}({\bf x}_j) + {\rm h.c.},
\end{align}
where $t^l_i$ is the hopping amplitude for $l=c, f$ electrons along $i$ direction,
and $\mu^l$ is the on-site chemical potential for $l=c, f$ electrons. 
The hopping Hamiltonian is explicitly invariant under $G_z$, $S_y$ and
$P$, and since all the other non-symmorphic symmetries can be expanded
in terms of these three (plus translation), the hopping is invariant
under the full non-symmorphic group. 

In momentum space, the hopping Hamiltonian is 
$H^{\rm hopping}=\sum_{\bf{k}} \Psi({\bf k})^\dagger \mathcal{H}^{\rm hopping}({\bf k}) \Psi({\bf k})$,
where 
where $\Psi\dg  ({\bf k})$ is a  sixteen component creation operator, $\Psi\dg ({\bf k})=(\Psi\dg _c({\bf k}),\Psi\dg _f({\bf k}))$ with
\begin{align} 
\Psi\dg _{c}({\bf k})=&(c\dg_{1A\uparrow }({\bf k}),c\dg_{1A\downarrow}({\bf k}),c\dg_{2A\uparrow }({\bf k}),c\dg_{2A\downarrow}({\bf k}), 
c\dg_{1B\uparrow }({\bf k}),c\dg_{1B\downarrow}({\bf k}),c\dg_{2B\uparrow }({\bf k}),c\dg_{2B\downarrow}({\bf k})),   \notag \\
\Psi\dg _{f}({\bf k})=&(f\dg_{1A\uparrow }({\bf k}),f\dg_{1A\downarrow}({\bf k}),f\dg_{2A\uparrow }({\bf k}),f\dg_{2A\downarrow}({\bf k}), 
f\dg_{1B\uparrow }({\bf k}),f\dg_{1B\downarrow}({\bf k}), f\dg_{2B\uparrow }({\bf k}),f\dg_{2B\downarrow}({\bf k})),   \notag
\end{align}
and
\begin{align}
\mathcal{H}^{\rm hopping}({\bf k})=\left(\begin{array}{cc}\mathcal{H}^{c}({\bf k}) & 0 \\0 & \mathcal{H}^{f}({\bf k})\end{array}\right),
\end{align}
with
\begin{align}
\mathcal{H}^{c/f} ({\bf k})=& (2 t^{c/f}_z \cos k_z - \mu^{c/f}) + 2 t^{c/f}_x \cos \frac{k_x}{2} ( \cos \frac{k_x}{2} \tau_1 + \sin \frac{k_x}{2} \tau_2 \rho_3) \notag\\
&+2 t^{c/f}_y \cos \frac{k_y}{2} ( \cos \frac{k_y}{2} \tau_1 \rho_1 + \sin \frac{k_y}{2} \tau_1 \rho_2). 
\label{Eq: hop}   
\end{align}

\subsubsection{Hybridization terms}
\begin{figure}[htbp]
\centering
\includegraphics[height=6cm] {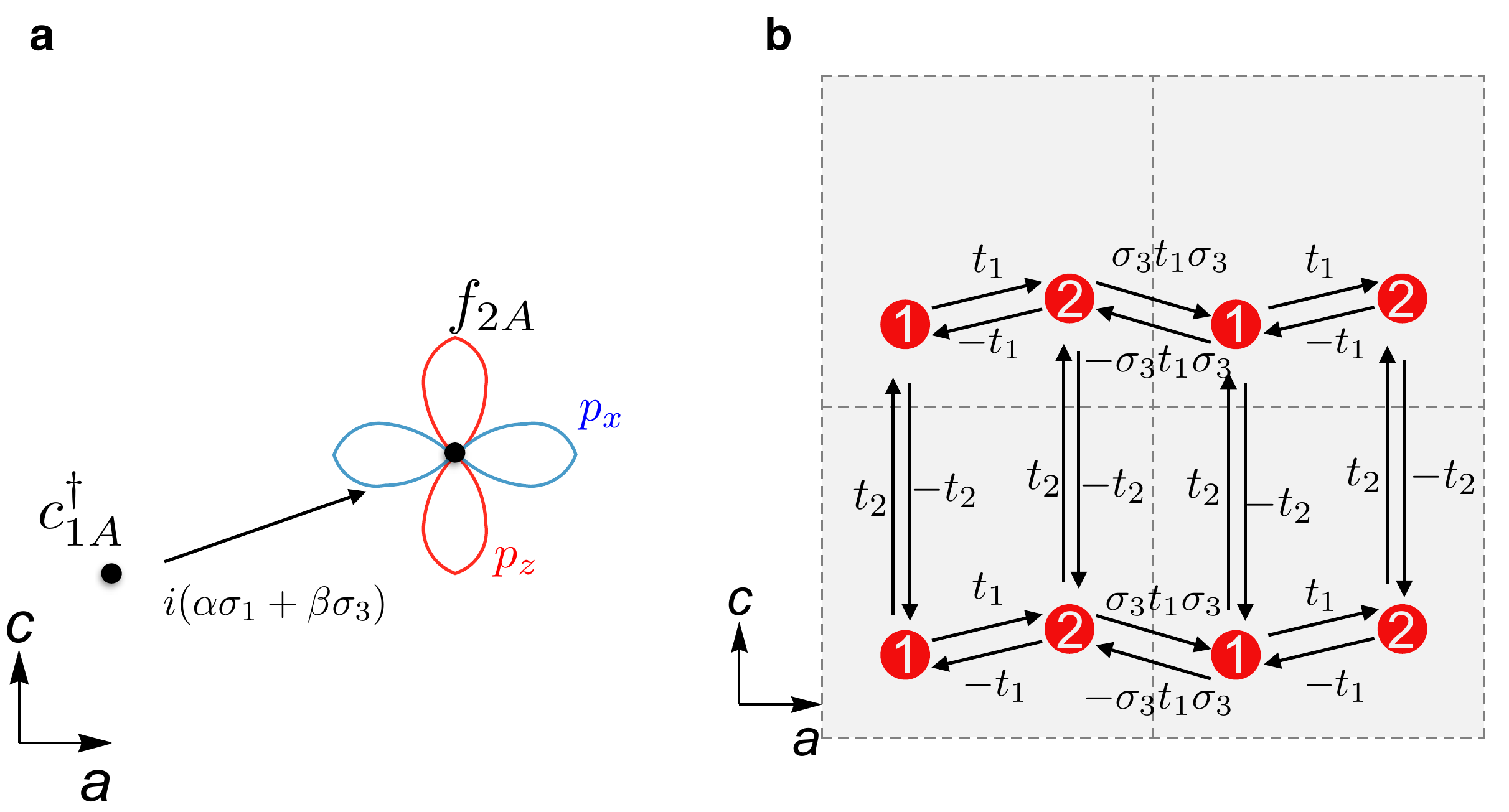}
\caption{{\bf Hybridization Hamiltonian on $A$ layer:}
{\bf a}, The hybridization matrix between $c^\dagger_{1A}$ and $f_{2B}$ is $t_1=i(\alpha \sigma_1+\beta \sigma_3)$.
{\bf b}, The hybridization elements $t_1$, $\sigma_3 t_1\sigma_3$ and $t_2$ on the $ac$ plane.}
\label{Fig: SM2}
\end{figure}
We now consider the hybridization terms between $f$-electrons and conduction electrons. 
Let us start with layer $A$.
The hybridization Hamiltonian on layer $A$ is (see Fig. \ref{Fig: SM2}{\bf b})
\begin{align}
H^{\rm hybr.}_A=& \sum_{j,\sigma,\sigma'} 
c^\dagger _{1A\sigma} ({\bf x}_{j}) [t_1]_{\sigma \sigma'} f_{2A\sigma'} ({\bf x}_{j})-
c^\dagger _{2A\sigma} ({\bf x}_{j}) [t_1]_{\sigma\sigma'} 
f_{1A\sigma'} ({\bf x}_{j})
+c^\dagger _{1A\sigma} ({\bf x}_{j}) 
[t_2]_{\sigma \sigma'} f_{1A\sigma'}({\bf x}_{j}+ c) \notag\\
&+c^\dagger _{2A\sigma} ({\bf x}_{j}) [t_2]_{\sigma \sigma'}
f_{2A\sigma'} ({\bf x}_{j}+c)  
+G_z^{-1}(\cdots) G_z +\mathrm{h.c.},
\end{align}
where $G_{z}^{-1}(\cdots)G_{z}$ in the second line is the 
glide-reflection of the
first four terms. This guarantees that ${G}_z^{-1}H^{\rm hybr.}_A {G}_z
=H^{\rm hybr.}_A$. 
The hybridization matrices have the structure 
$t_{1} = i (\alpha \sigma_1+\beta \sigma_3)$, and $t_{2} = i \gamma \sigma_3$ (see Fig. \ref{Fig: SM2}{\bf a}). 
In momentum space, the hybridization Hamiltonian on layer $A$ 
is $H^{\rm hybr.}_A=\sum_{\bf{k}} \Psi\dg _A({\bf k}) \mathcal{H}_A^{\rm hybr.}({\bf k}) \Psi_A({\bf k})$,
where 
\begin{equation}\label{}
 \Psi\dg  _A({\bf k})=(c\dg_{1A\uparrow }({\bf k}),c\dg_{1A\downarrow}({\bf k}),c\dg_{2A\uparrow }({\bf k}),c\dg_{2A\downarrow}({\bf k}),
f\dg_{1A\uparrow }({\bf k}),f\dg_{1A\downarrow}({\bf k}),f\dg_{2A\uparrow }({\bf k}),f\dg_{2A\downarrow}({\bf k}))
\end{equation} 
and
\begin{align}
\mathcal{H}^{\rm hybr.}_A({\bf k})=\left(\begin{array}{cc}0 & V_A({\bf k}) \\ V_A({\bf k})^\dagger & 0 \end{array}\right)
,\quad \mathrm{with} \quad
V_A({\bf k})  
=
\left(\begin{array}{cc}  2 i t_2 \sin k_z & t_1+\sigma_3 t_1 \sigma_3 e^{-i k_x} 
\\ -t_1-\sigma_3 t_1 \sigma_3 e^{i k_x} & 2i t_2 \sin k_z 
\end{array}\right).
\label{Eq: VA}
\end{align}

The hybridization Hamiltonian on the $B$ layer can be obtained by performing
the screw rotation $S_y$ on $H^{\rm hybr.}_A$, $H^{\rm hybr.}_B=S^{-1}_y H^{\rm hybr.}_A S_y$.

\begin{figure}[htbp]
\centering
\includegraphics[height=5cm] {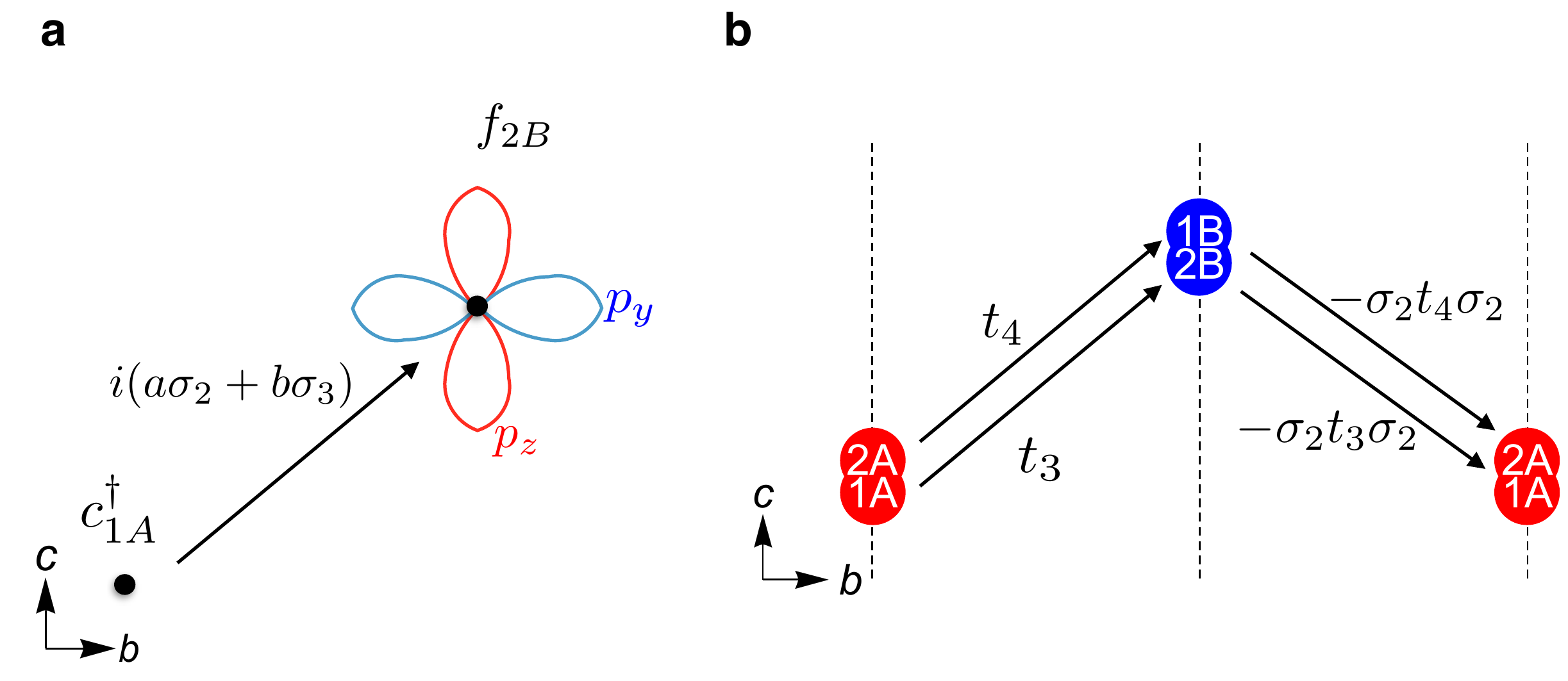}
\caption{{\bf Interlayer hybridization Hamiltonian:}
{\bf a}, The interlayer hybridization matrix between $c^\dagger_{1A}$ and $f_{2A}$ is 
$t_3=i(a \sigma_2+b \sigma_3)$.
{\bf b}, The interlayer hybridization elements $t_3$ and $t_4$ on the $bc$ plane.}
\label{Fig: SM3}
\end{figure}
Next, we write down the interlayer hybridization terms.
The interlayer hybridization Hamiltonian (see Fig. \ref{Fig: SM3}{\bf b}) is
\begin{align}
H_{AB}^{\rm hybr.} =& \sum_{j, \sigma, \sigma'}
c^\dagger _{1A\sigma}({\bf x}_j) [t_3]_{\sigma \sigma'} f_{2B\sigma'}({\bf x}_j)+
c^\dagger _{2A\sigma}({\bf x}_j) [t_4]_{\sigma \sigma'} f_{1B\sigma'}({\bf x}_j)+  \notag\\
&+ {S}_y^{-1} (\cdots) {S}_y+ {P}^{-1} (\cdots) {P}
+ [{P}{S}_y]^{-1} (\cdots) {P}{S}_y + \mathrm{h.c.},
\end{align}
where $ (\cdots) $ denote a repeat of the first two terms, 
$t_3= i (a \sigma_2 +b \sigma_3)$ and $t_4= i (a \sigma_2 -b
\sigma_3)$ (see Fig. \ref{Fig: SM3}{\bf a}).
By construction,  interlayer hybridization is invariant under $G_{z}$
$S_{y}$ and $P$, and is thus invariant under the full non-symmorphic group
In momentum space, the interlayer hybridization Hamiltonian is
$H^{\rm hybr.}_{AB}=\sum_{\bf{k}} \Psi({\bf k})^\dagger \mathcal{H}_{AB}^{\rm hybr.}({\bf k}) \Psi({\bf k})$,
where
\begin{align}
&\mathcal{H}_{AB}^{\rm hybr.}({\bf k})
=\left(\begin{array}{cccc} &  &  & V_{AB}^{} ({\bf k}) \\ &  &V_{BA}^{} ({\bf k}) ^\dagger  &  
\\ & V_{BA}^{} ({\bf k}) &  &  \\  V_{AB}^{} ({\bf k})^\dagger  &  &  & \end{array}\right),  \quad \mathrm{with} \notag\\
&V_{AB}^{} ({\bf k}) =
\left(\begin{array}{cc} &t_3-\sigma_2 t_3 \sigma_2 e^{-ik_y} 
 \\  t_4- \sigma_2 t_4 \sigma_2e^{-ik_y}  &\end{array}\right), \quad
V_{BA}^{} ({\bf k}) =
\left(\begin{array}{cc}  & \sigma_2 t_3 \sigma_2 e^{ik_y} -  t_3 
\\ \sigma_2 t_4 \sigma_2 e^{ik_y}- t_4 & \end{array}\right).
\label{Eq: VAB}
\end{align}

\subsubsection{Full Hamitonian}\label{}

Finally, we obtain the  
total single-particle Hamiltonian
\begin{align}
&\mathcal{H}^{}({\bf k})=
\left(\begin{array}{cc} \mathcal{H}^c({\bf k}) & V({\bf k}) \\ V^\dagger({\bf k}) & \mathcal{H}^f({\bf k})\end{array}\right),  
\end{align}
where $ \mathcal{H}^{c/f}({\bf k})$ is defined in equation (\ref{Eq: hop}),
and 
\begin{align}
V({\bf k})=
\left(\begin{array}{cc} V_A({\bf k}) & V_{AB}({\bf k}) \\ V_{BA}({\bf k}) & V_B({\bf k})\end{array}\right),    \notag
\end{align}
with $V_A({\bf k})$ being defined in equation (\ref{Eq: VA})
and $V_{AB}({\bf k})$, $V_{BA}({\bf k})$ being defined in equation (\ref{Eq: VAB}).

\subsubsection{Matrix representation of symmetries}
Once we fix the basis of the spinor $\Psi ({\bf k})$, we can write down all matrix representations
of symmetries in momentum space.
The glide reflection is
\begin{align}
{\mathcal{G}_z}({\bf k}) 
=-ie^{-i\frac{k_x}{2}}\sigma_3(\cos \frac{k_x}{2} \tau_1 + \sin \frac{k_x}{2} \tau_2 \rho_3) \lambda_3,
\end{align}
while the screw rotation and inversion symmetry in the momentum space are
\begin{align}
&{\mathcal{S}_y}({\bf k}) 
=-ie^{-i\frac{k_y}{2}}\sigma_2\tau_0\lambda_0(\cos \frac{k_y}{2} \rho_1+\sin \frac{k_y}{2} \rho_2),    \notag \\
&{\mathcal{P}} 
=\sigma_0\tau_0  \lambda_3 \rho_1.
\end{align}
It can be verified that 
${\mathcal{G}_z}({\bf k})^{-1} \mathcal{H}(k_x,k_y,k_z) {\mathcal{G}_z}({\bf k}) =  
\mathcal{H}(k_x,k_y,-k_z)$, and that 
${\mathcal{G}_z}({\bf k})^2 = -e^{ - i k_x}$, as expected from the
combination of full translation and twice reflections
(see equation (\ref{Eq: G2}) in main text).
Similarly, one can verify that the Hamiltonian transforms under
$\mathcal{S}_{y}$ as 
$\mathcal{S}_y({\bf k})^{-1} \mathcal{H}(k_x,k_y,k_z) \mathcal{S}_y({\bf k}) =  
\mathcal{H}(-k_x,k_y,-k_z)$ and under inversion, 
${\mathcal{P}}^{-1} \mathcal{H}({\bf k})\mathcal{P} =  
\mathcal{H}(-{\bf k})$.
In a similar fashion to the glide reflection, ${\mathcal{S}_y}({\bf k})^2 = -e^{ - i k_y}$,
which is expected from a $2 \pi$ rotation and one full lattice
translation. 
The transformation of the Hamiltonian under the full set of
non-symmorphic symmetries can be obtained by decomposing them in terms
of $G_{z}$ and $S_{y}$, as described in (\ref{othersym}).

Time-reversal symmetry has the usual representation, $\mathcal{T}= i \sigma_2 \mathcal{K}$,
where $\mathcal{K}$ is the complex conjugation operator.
We have $\mathcal{T}^{-1} \mathcal{H}({\bf k})\mathcal{T} = \mathcal{H}(-{\bf k})$.

\subsection{$ \mathbb{Z}_4$ topological invariant}
\label{App: Z4}

\begin{figure}[htbp]
\centering
\includegraphics[height=4cm] {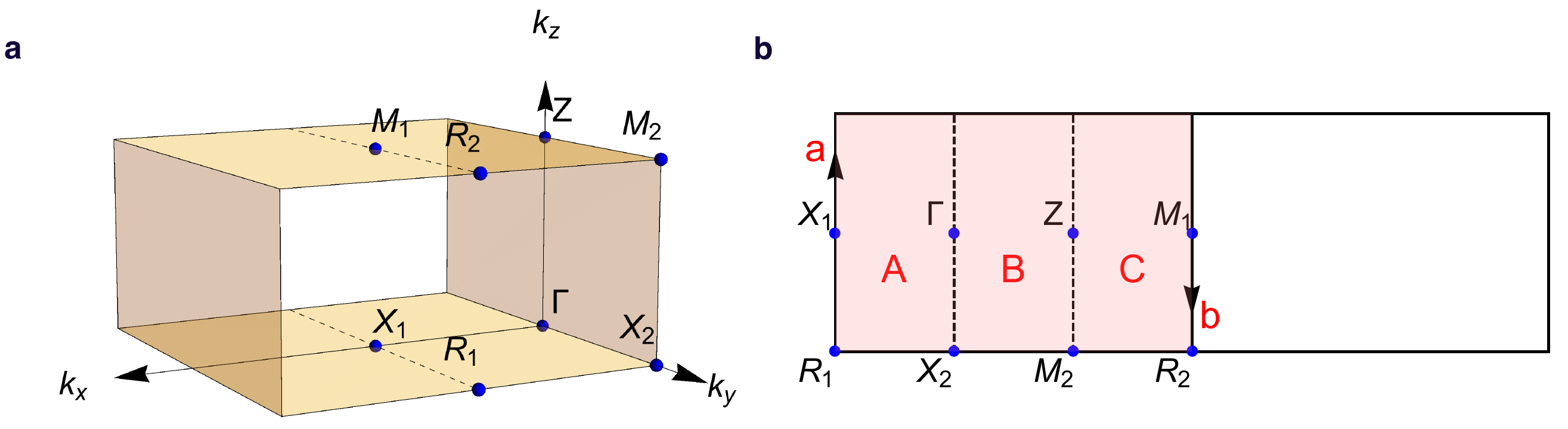}
\caption{{\bf Illustration of calculating the $\mathbb{Z}_4$ invariant.}
{\bf a}, The "bent" BZ (shaded in yellow) is chosen by connecting two glide planes to their neighboring plane,
which is traversing half the BZ at $k_x=0$.
{\bf b}, The  $\mathbb{Z}_4$ invariant in equation (\ref{Eq: Z4}) is calculated in the half of the "bent" BZ
(shaded in red).}
\label{Fig: Z4}
\end{figure}

We give the definition of the $\mathbb{Z}_4$ invariant in symmetry class AII introduced in Ref. \cite{Shiozaki2015}.
In the presence of the glide reflection symmetry $\mathcal{G}_z$,
there are two glide planes in the BZ at $k_z=0$ and $k_z=\pi$. In these glide planes,
the occupied states can be separated into two sectors with the glide eigenvalues $g_{\pm}=\pm i e^{-i k_x/2}$.
Due to the bulk-boundary correspondence, the existence of protected
surface modes on glide planes are associated with the 
Berry connections and Berry curvature on the "bent" BZ \cite{Alexandradinata}.
This "bent" BZ is chosen by connecting two glide planes to their neighboring plane,
which is traversing half the BZ at $k_x=0$ (Fig. \ref{Fig: Z4}{\bf a}).
In the main text, we demonstrate the number of Dirac cones
along path $X'\Gamma X M Z M' X'$ is modulo four,
which implies that the $\mathbb{Z}_4$ invariant
can be defined from calculating the winding number of the Berry connections
of two glide sectors on the "bent" BZ.
Having the same spirit of calculating the $\mathbb{Z}_2$ invariant in time-reversal symmetric system,
we only need to consider the half of the "bent" BZ as shown in Fig. \ref{Fig: Z4}{\bf b}.
Along paths $a$ and $b$, the eigenvalues of glide reflection symmetry are real and
the Kramers pairs $(|u_{\mu}^{\pm, I}  ({\bf k}) \rangle, |u_{\mu}^{\pm, II}  ({\bf k}) \rangle)$ are in the same glide sector.
We have
\begin{align}
\oint_{a (b)} d l \mathcal{A}^{\pm} =  2 \oint_{a (b)} d l \mathcal{A}^{\pm,I} \quad \mathrm{mod}  \quad 2\pi,
\end{align}
where  $\mathcal{A}^{\pm,I} =i \sum_{\mu \in \mathrm{occ.}} \langle u_{\mu}^{\pm,I} ({\bf k}) | \partial_{k_y} u_{\mu}^{\pm,I}  ({\bf k}) \rangle$,
and  $\mathcal{A}^{\pm} =i \sum_{\mu \in \mathrm{occ.}} \langle u_{\mu}^{\pm,I} ({\bf k}) | \partial_{k_y} u_{\mu}^{\pm,I}  ({\bf k}) \rangle
+ \langle u_{\mu}^{\pm,II} ({\bf k}) | \partial_{k_y} u_{\mu}^{\pm,II}  ({\bf k}) \rangle$.

The $\mathbb{Z}_4$ invariant is defined by consider the positive glide sector at glide plane. 
We have
\begin{align}
\chi:= \frac{1}{2 \pi} [4\oint_{a} d l \mathcal{A}^{+,I}  - 4\oint_{b} d l \mathcal{A}^{+,I}
-(2 \int_{A} d a \mathcal{F}^+ + 2 \int_C da \mathcal{F}^+ + \int_B da \mathcal{F})],  \quad \mathrm{mod} \quad 4,
\label{Eq: Z4}
\end{align}
where the Berry curvature is defined as $\mathcal{F}^{\pm} = \partial_t \mathcal{A}^{\pm}-\partial_{k_y} \mathcal{A}^{\pm}$
with $t$ being the momentum direction which perpendicular to $k_y$ on the "bent BZ"
and $\mathcal{F} = \mathcal{F}^{+}+\mathcal{F}^{-}$.

\subsection{Resistivity plateau in doped C\lowercase{e}N\lowercase{i}$_{1-\delta}$Sn$_{1+\delta-x}$S\lowercase{b}$_x$  and C\lowercase{e}$_3$B\lowercase{i}$_4$P\lowercase{t}$_3$ }

\begin{figure}[htbp]
\centering
\includegraphics[height=6cm] {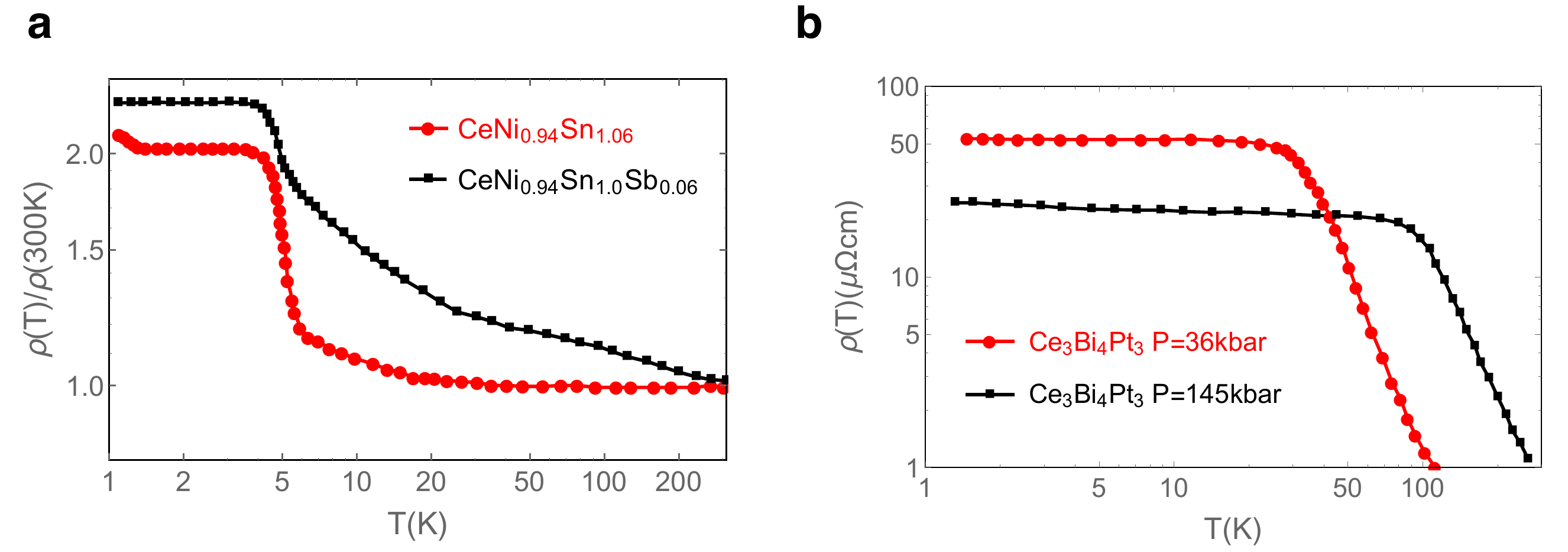}
\caption{{\bf Resistivity versus temperature adapted from Refs. [\onlinecite{Cooley_PRB97, SlebarskiX}]:}
{\bf a}, doped CeNi$_{1-\delta}$Sn$_{1+\delta-x}$Sb$_x$.
{\bf b}, Ce$_3$Bi$_4$Pt$_3$ under pressure.}
\label{Fig: SM4}
\end{figure}

Fig \ref{Fig: SM4} shows the resistivity plateau at low temperature of doped CeNi$_{1-\delta}$Sn$_{1+\delta-x}$Sb$_x$  and Ce$_3$Bi$_4$Pt$_3$ under pressure.

\bibliographystyle{naturemag}

\end{document}